# Switchable polar spirals in tricolor oxide superlattices

*Zijian Hong* and *Long-Qing Chen*

Department of Materials Science and Engineering, The Pennsylvania State University, University Park, PA 16802, USA

**ABSTRACT**: There are increasing evidences that ferroelectric states at the nanoscale can exhibit fascinating topological structures including polar vortices and skyrmions, akin to those observed in the ferromagnetic systems. Here we report the discovery of a new type of polar topological structure–ordered array of nanoscale spirals in the $PbTiO_3/BiFeO_3/SrTiO_3$ tricolor ferroelectric superlattice system via phase-field simulations. This polar spiral structure is composed of fine ordered semi-vortex arrays with vortex cores forming a wavy distribution. It is demonstrated that this tricolor system has an ultrahigh Curie temperature of ~1000 K and a temperature of ~650 K for the phase transformation from spiral structure to in-plane orthorhombic domain structure, showing a greatly enhanced thermal stability than the recently discovered polar vortex lattices in the $PbTiO_3/SrTiO_3$ superlattice system. Moreover, the spiral structure has a net in-plane polarization that could be switched by an experimentally-feasible irrotational in-plane field. The switching process involves a metastable vortex state, and is fully reversible. This discovery could open up a new routine to design novel polar topological structures with enhanced stability and tunability towards future applications in next-generation nanoscale electronics.

**KEYWORDS**: polar spiral, thermal stability, switching dynamics, phase-field

simulations, nanoscale electronics

Nanoscale topological structures and their phase transitions in the ferroic materials have received great attention due to the fact that they are not only scientifically fascinating but also have potential applications in electronic devices such as memories and logic gates. For instance, vortices[1-6], skyrmions[7-9] and merons[10, 11] etc. were discovered in the past few decades in both ferroelectric and ferromagnetic materials. They can be manipulated by external stimuli such as magnetic/electric field or electric current. It is demonstrated that one can move and switch ferromagnetic vortices and skyrmions using an external electric current[12], and their device applications have been proposed[13]. Compared to the conventional spintronic devices, these novel topological structures with greatly reduced sizes could facilitate the miniaturization of next-generation electronic and spintronic devices. However, their thermodynamic stability as well as the mesoscale mechanisms of their transformations from one type of structure to another are not yet well understood.

One exciting recent advance in polar topological structures and phase transitions is the discovery of the nanoscale ferroelectric vortex arrays in $(PbTiO_3)_n/(SrTiO_3)_n$ ($n$=10-16) (PTO/STO) superlattices on a $DyScO_3$ (DSO) substrate[14, 15]. While such vortex arrays are scientifically interesting, there are at least two main obstacles to realizing their device applications. Firstly, it is not trivial to switch the curl of the polar vortices by means of an irrotational electric field. Theoretical studies have demonstrated that the curl of a vortex in ferroelectrics can be switched either with a careful design of the device geometry[16-18] or by applying an inhomogeneous or a curled electric field[19, 20], which however, is experimentally challenging or even unfeasible. Secondly, the polar vortex lattice in this system is thermally unstable, favoring the formation of $a_1/a_2$ twin domain structures upon heating. Recent experimental and theoretical studies have shown that the vortex lattice transforms to $a_1/a_2$ twin domain structure gradually with increasing temperature, vanishing close to ~500 K[21].

BiFeO$_3$ (BFO) has long been considered as one of the most promising room temperature multiferroic materials (with both room temperature ferroelectric and G-type antiferromagnetic order), which is under extensive investigation in the past decade [22-25]. It has a much higher Curie temperature than other prototype ferroelectric materials (~1100 K[23], compared with PTO with Curie temperature of ~750 K and BaTiO$_3$ with Curie temperature of 400 K). At room temperature, bulk BFO has a large spontaneous polarization (~100 μC/cm$^2$) with a space group of $R3c$[23]. So far, to the best of our knowledge, only few vortex-like or flux-closure structures have been observed in BFO-based heterogeneous thin films or superlattices[6, 26-28] since it is difficult for polarization to form continuously rotating patterns due to the relative "rigid" nature of the polarization directions or the strong anisotropy in BFO.

Here we consider a PTO/BFO/STO tricolor model system (hereafter referred as PBS-tricolor) in which the repeating unit consists of 4 unit cells of BFO sandwiched between two blocks of PTO layers (4 unit cells in each block), followed by 12 unit cells of insulating paraelectric STO layers (see the schematics of the building blocks in Figure 1). The whole film is fully strained to a $(110)_o$-DSO substrate (the lattice constants of substrate, PTO, BFO and STO are given in the supplementary to determine the strain conditions in each layer). In comparison, 12 unit cells of PTO and 12 unit cells of STO are periodically stacked to form a $(PTO)_{12}/(STO)_{12}$ superlattice (referred as PST-superlattice, schematics shown in Figure S1). The PBS-tricolor system can be regarded as the PST-superlattice with the middle PTO layers substituted by BFO layers.

Phase-field simulations are performed by solving the time-dependent Ginzburg-Landau equations for the spatial distribution of spontaneous polarization $\vec{P}$:

$$\frac{\partial \vec{P}}{\partial t} = -L \frac{\delta F}{\delta \vec{P}} \quad \text{(Equation 1)}$$

where $t$ and $L$ represent the evolution time and kinetic coefficient, respectively. $F$ is the total free energy of the system including contributions from elastic, electric, Landau/chemical and polarization gradient energies:

$$F = \int (f_{Elas} + f_{Elec} + f_{Land} + f_{Grad}) dV \quad \text{(Equation 2)}$$

Detailed descriptions of solving the phase-field equations can be found elsewhere[15, 29-31]. Thermordynamic potentials as well as other material constants are adopted from previous reports[32-36]. The simulation system is discretized into a three dimensional mesh of 200×200×250, with each grid representing 0.4 nm. The thickness direction consists of 30 numerical grids of substrate, 192 grids of film, and 28 grids of air, respectively. Periodic boundary condition is applied along the in-plane dimensions while the thickness direction is solved using a superposition method[37]. To obtain the electrostatic energy contribution to the polarization state, the short-circuit electric boundary condition is used by fixing the electric potential at the top and bottom of the film to 0. To perform in-plane field switching, a cyclinic homogenous in-plane electric field is applied to the film until it reaches 400 kV/cm in magnitude. For the mechanical energy contribution, a thin film mechanical boundary condition is applied with a stress free condition on the film surface and zero displacements within the substrate sufficiently far away from the film/substrate interface[30]. The iteration pertubation method is adopted to solve the elastic equlibrium equation taking into account the differences in the elastic constants among the PTO, BFO and STO layers[38].

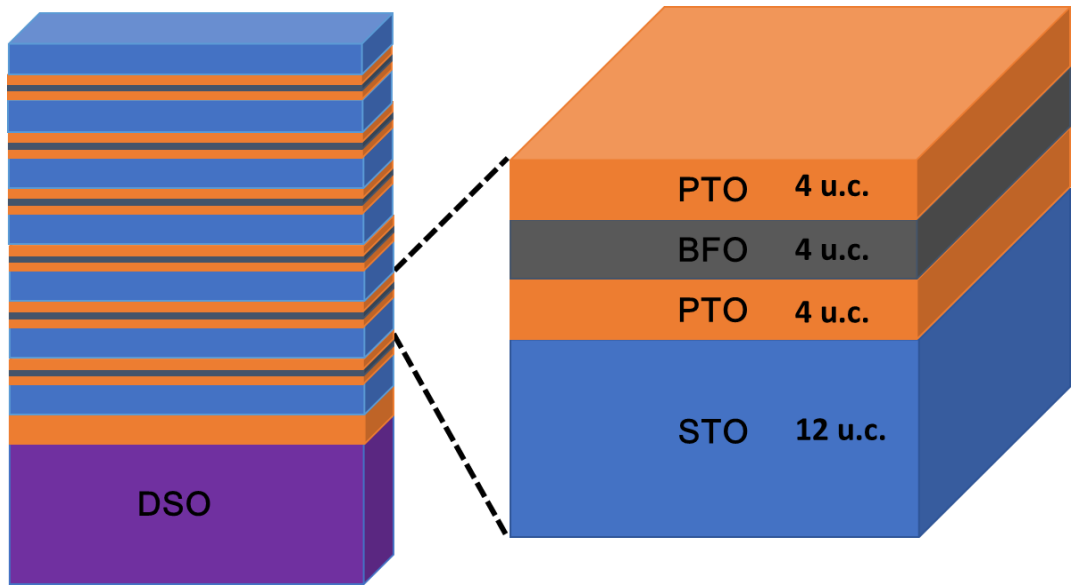

**Figure 1.** Schematics of the model superlattice system. Left, the whole simulation cell; right, the single periodic unit, consists of PTO, BFO, PTO, each 4 unit cells and 12 unit cells of STO.

As expected, the PST-superlattice at room temperature exhibits an ordered arrays of vortices with long tube-like vortex lines, where the vortex cores are close to the center of PTO layers (Figure S2). The formation of vortex arrays within the PST-superlattice has previously been analyzed in details[14, 15]. Upon substitution of the middle PTO layers with BFO, one intuitive question is whether the vortex structure could be stabilized. To answer this question, the polar structure of PBS-tricolor system at room temperature is simulated and plotted in Figure 2. The planar view image in Figure 2a shows that periodic long stripes form, similar to the vortex lines in the PST-superlattice. However, these stripes are highly curved, forming a wave like pattern. The zoom-in plot in Figure 2b indicates that a large $P_y$-component is found in the PTO layers, which causes the wave-like vortex lines. The $P_y$-component is largely induced by the rhombohedral BFO layer, leading to the large polarization rotation in PTO layers to account for the large symmetry mismatch between PTO and BFO layers.

The cross-section view image (Figure 2c) demonstrates that a unidirectional polar spiral structure forms, with significantly larger polarizations in the middle BFO layers. This is in contrast to the vortex lattice structure in PST-superlattices where polarization in the cores near the middle of the PTO layers is largely reduced (see Figure S2). This can be understood since the bulk BFO has a large spontaneous polarization, and reducing the polarization in BFO layers would lead to a significant increase in the chemical Landau energy. A magnified view in Figure 2d clearly shows that the spirals are composed of ordered arrays of alternating semi-vortices in different layers, with vortex cores floating up-down with respect to the middle BFO layers, also forming a wave-like pattern with even smaller length scale. Due to the strong interfacial and electromechanical coupling, both the polarization in PTO and BFO layers are distorted from the corresponding bulk tetragonal and rhombohedral directions. These distortions reduce the polar discontinuity at two interfaces, thus minimize the surface charges as well as the electric energy. The periodicity of a spiral is ~10 nm, which is close to the size of two vortices in the PST-superlattice. Here, it is demonstrated that the introduction of middle BFO layers with larger polarization and a symmetry mismatch between BFO and PTO layers could lead to the formation of a unidirectional spiral structure.

In a direct comparison, the PST-superlattice always involves the mutually counter-rotations of polarization in the neighboring vortices with minimal net polarization, giving rise to very small piezo-/dielectric- responses; whereas a spiral structure in our newly designed PBS-tricolor system has a relatively large net in-plane polarization, which could potentially give large in-plane PFM signal, facilitating the characterization and ultimately the future applications of this phase.

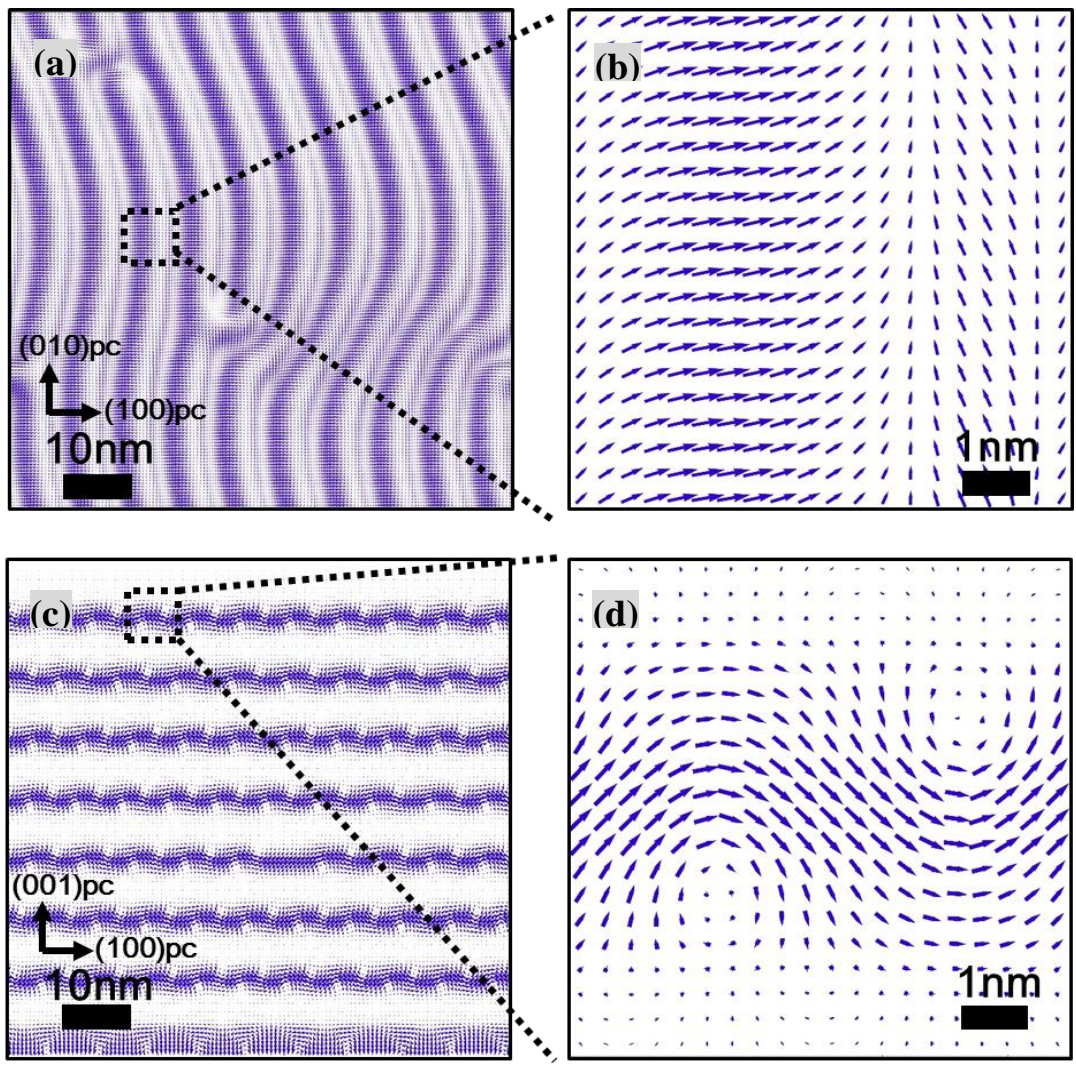

**Figure 2.** Polar mapping of the tri-color system. (a) Polarization vector of the planar view in the PTO layer, showing the formation of curved stripes (b) Magnified planar view, a large y-component is clearly shown close to the main stripe. (c) Cross-section view, showing that the polar vector is forming unidirectional spirals in each layer. (d) Magnified cross-section view, showing the spirals consist of two half-vortices.

In order to reveal the thermal stability of a polar spiral phase, the temperature phase diagrams for both PBS-tricolor and PST-superlattice systems have been simulated and compared in Figure 3. The mean square of the total polarization (defined as $\langle P^2 \rangle = \frac{\int (P_1^2 + P_2^2 + P_3^2) dV}{V}$) is plotted with respect to temperature, which shows a linear decay in both two materials with similar slopes until reaching zero. The calculated Curie temperatures can be extracted where the linear lines intersect with zero polarization. It is discovered that the Curie temperature shows a large decrease in a PST-superlattice (~650 K) as compared to bulk PTO (~750 K), due to the large depolarization field. The decrease in Curie temperature with the reduction in the size of ferroelectrics in the ferroelectric/paraelectric superlattices as well as ultrathin films has been well studied both theoretically[39, 40] and experimentally[41, 42]. For the PST-superlattice, the Curie temperature obtained here is in good agreement with the experimentally measured value for PTO thin films grown on a $(110)_o$-DSO substrate with a large depolarization field[43]. In a PBS-tricolor system, the calculated Curie temperature has a huge increase (~1000 K) as compared to a PST-superlattice system, which is even higher than the Curie temperature of bulk PTO. This can be understood from two aspects: Firstly, BFO has a high Curie temperature (~1100 K), and at the temperature range between 650 K-1000 K, even though polar PTO is unstable, polar BFO is energetically preferred, which could increase the Curie temperature of the whole system; Secondly, a large internal field imposed by polar BFO layers could serve as a self-poling field, which ultimately increases the Curie temperature of PTO layers. For example, a polar mapping of the high temperature phase (see Figure S3) shows that the ferroelectric BFO layers are able to polarize the PTO layers near the BFO/PTO interfaces.

To get a better insight of the phase transformations for the PBS-tricolor below Curie temperature (~1000K), the mean square of the out-of-plane polarizations (defined as $\langle P_3^2 \rangle = \frac{\int P_3^2 dV}{V}$) is plotted as a function of temperature (see inset of Figure 3). Upon increasing the temperature, the mean square of out-of-plane polarization decreases until reaching zero at ~650K, indicating the formation of in-plane domains beyond this temperature. This transition point almost coincides with the Curie temperature of PST-superlattice, which can be understood since polar PTO phase is less stable above this temperature, and the out-of-plane polarization in BFO layer will induce large polar discontinuity and hence the depolarization field, which is energetically unfavorable. As a result, a transformation to in-plane domain occurs. Further investigation reveals that this in-plane domain state is an orthorhombic twin structure (Figure S4). Previously, the strain-temperature phase diagram obtained by phase-field simulations indicates that the orthorhombic phase can be stabilized at moderate strains with the open-circuit electric boundary condition at relatively high temperature, while experimental studies have indeed observed the orthorhombic BFO phase under a tensile substrate strain[44].

In a direct comparison, the complete transformation of vortex to $a_1/a_2$ in the PST-superlattice occurs at a much lower temperature (~500 K). One can conclude that the PBS-tricolor system with the addition of BFO layers exhibits both increased Curie temperature and topological to non-topological phase transformation temperature, largely arising from the large polarization and high Curie temperature of BFO.

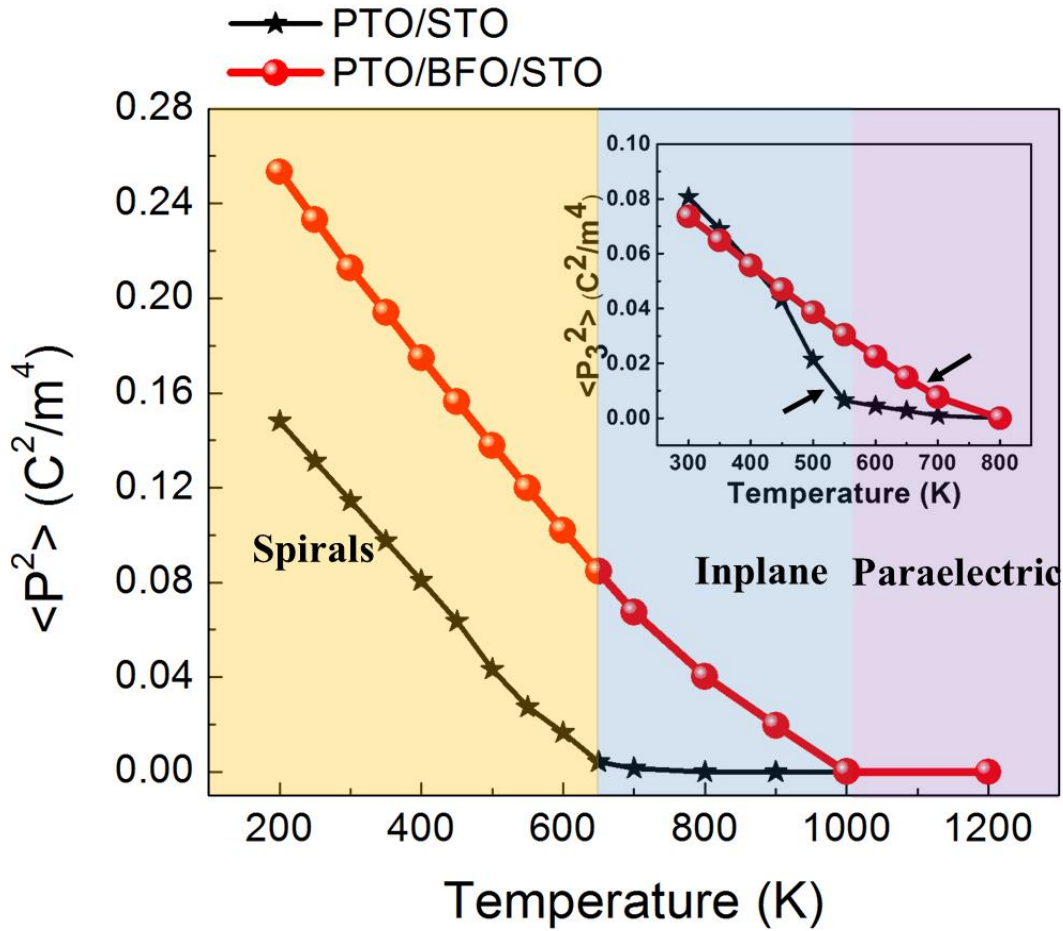

**Figure 3**. Temperature phase diagram of the tri-color system and a comparison with the PTO/STO superlattice. Huge increase of Curie temperature can be observed. Inset: Stability of the out-of-plane polarization, which indicate the transition between spirals (or vortex) and inplane domains. Arrows mark the transition temperature, which shows a large upshift.

One unique feature of the spiral structure in a PBS-tricolor system as compared to the polar vortex phase in a PST-superlattice system is that it possesses a net in-plane polarization. One natural question towards the potential applications of this novel structure would be: can we switch the net in-plane polarization direction of the spirals by an irrotational field. The electric field switching process is simulated by applying a uniform in-plane electric field (Figure 4). Initially, without an external bias, the spirals are curving to the right with a net +$P_x$, consisting of periodic semi-vortices that are floating up-down. An in-plane electric field with a magnitude of 350 kV/cm is then applied, which is opposite to the initial spiral direction. This field could ultimately lead to the switching of in-plane polarization component to -$P_x$ inside the PTO layers, while the in-plane components in BFO layers are not yet switched (Figure 4b). As a result, ordered vortex-like array structure similar to a PST-superlattice is formed. It should be pointed out that the field-induced metastable vortex-like structure in the PBS system is not fully circular due to the difficulty in rotating the polarizations in BFO layers (in other word, BFO is more "stiff"). At even higher fields (e.g., ~400 kV/cm), the in-plane polarization of the BFO layers switches, thus switching the direction of the spirals (Figure 4c). This structure could be stabilized even when the field is removed (Figure 4d). Further switching studies indicate that the whole process is fully reversible; with the application of a positive in-plane field, the direction of spirals can be switched back to +$P_x$ again and stabilized after the applied field is removed. Here, it is shown that the directions of the spirals could be switched back and forth by experimentally feasible in-plane electric fields. Also, it should be mentioned that the reversible switching process involves multiple distinguishable states (spirals, vortex and possibly even pure *a*-domains), which could potentially be explored for applications (e.g., multiple state memory devices[45], logic gates[46], neuromorphic computing[47], etc.).

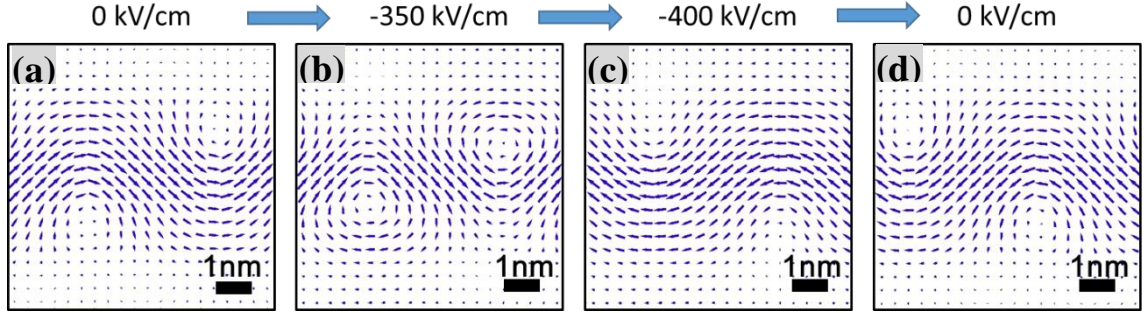

**Figure 4**. Switching of the spiral direction with in-plane field that is opposite to the initial spiral direction.

To conclude, we have simulated the polar structures of the PTO/BFO/STO system and studied its thermal stability and switching kinetics using phase-field simulations. It is revealed that a spiral phase is formed at room temperature with semi-vortex cores floating up and down with a wave-like feature, giving rise to a net in-plane polarization. This is a reminiscent of the polar vortex array structure that has been discovered in the PST-superlattice system[14, 15]. The PBS-tricolor system shows greatly increased Curie temperature and enhanced thermal stability for the spiral phase as compared to the vortex lattice in PST-superlattice system with the substitution of some PTO layers by the high Curie temperature BFO layers. The spiral to in-plane orthorhombic domain transition temperature is even close to the Curie temperature of the PST-superlattice (~650 K), and is much higher than the transition temperature of vortex to $a_1/a_2$ in the PST-superlattice. Further simulation results show that the spiral structure can be reversibly switched by experimental accessible in-plane electric fields, which involves a metastable vortex structure in-between two spiral phases with opposite in-plane direction.

## ASSOCIATED CONTENT

**Supporting Information**. Detailed description of the phase-field methodology, polar vortex in the PST-superlattice, high temperature domain structure in the PBS-tricolor system, planar view of the in-plane switching.

## AUTHOR INFORMATION

**Corresponding Author**

*E-mail: zxh121@psu.edu; lqc3@psu.edu.

**Acknowledgements:**

The work is supported by the U.S. DOE, Office of Basic Energy Sciences, Division of Materials Sciences and Engineering under Award number DE-FG02-07ER46417 (ZJH and LQC) and the NSF-MRSEC grant number DMR-1420620 and NSF-MWN grant number DMR-1210588 (ZJH). ZJH would like to thank Dr. R. Ramesh for useful comments and suggestions.

Table of Contents Graphic (For Table of Contents Only)

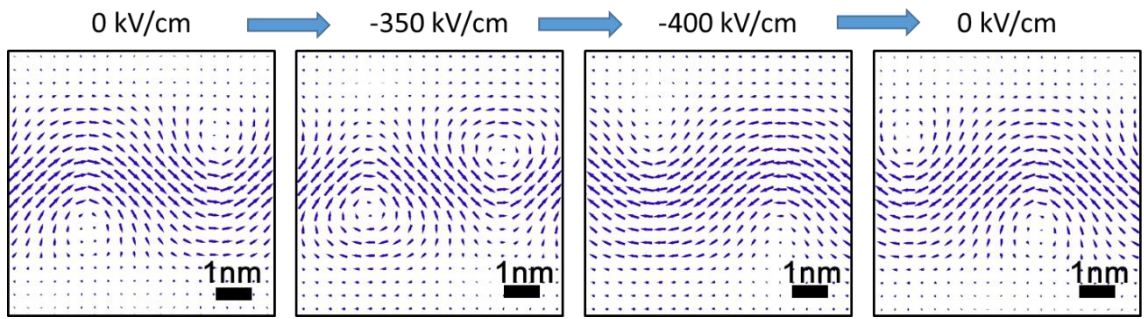